\documentclass[prd,showpacs,twocolumn,superscriptaddress,nofootinbib,,floatfix,10pt]{revtex4-2}
\usepackage{setspace}
\usepackage[utf8]{inputenc}
\usepackage{float}
\usepackage{amsmath,amssymb,amsfonts,bm}
\usepackage{graphicx}
\usepackage{multirow}
\usepackage{slashed}
\usepackage[usenames,dvipsnames]{color}
\allowdisplaybreaks
\usepackage[
colorlinks=true,
linkcolor=blue,
breaklinks=true,
urlcolor=blue,
citecolor=blue]{hyperref}

\usepackage{orcidlink}

\definecolor{green}{rgb}{0,0.6,0}

\newcommand{\ket}[1]{\left| #1 \right\rangle}

\newcommand{\be}{\begin{equation}} 
\newcommand{\ee}{\end{equation}}
\newcommand{\bea}{\begin{eqnarray}} 
\newcommand{\eea}{\end{eqnarray}}
\newcommand{\beas}{\begin{eqnarray*}} 
\newcommand{\eeas}{\end{eqnarray*}}

\newcommand{\bonn}{\affiliation{Helmholtz-Institut f\"ur Strahlen- und Kernphysik and Bethe Center for Theoretical Physics,\\ Universit\"at Bonn, D-53115 Bonn, Germany}}

\newcommand{\fzj}{\affiliation{Institute for Advanced Simulation, Institut f\"ur Kernphysik and J\"ulich Center for Hadron Physics,\\ Forschungszentrum J\"ulich, D-52425 J\"ulich, Germany}}

\newcommand{\Tbilisi}{\affiliation{Tbilisi State University, 0186 Tbilisi, Georgia}}

\begin{document}
\author{Teng Ji\orcidlink{0000-0003-0366-1042}}
\email{teng@hiskp.uni-bonn.de}\bonn

\author{Xiang-Kun Dong\orcidlink{0000-0001-6392-7143}}\email{xiangkun@hiskp.uni-bonn.de}\bonn

\author{Ulf-G. Mei{\ss}ner\orcidlink{0000-0003-1254-442X}}\email{meissner@hiskp.uni-bonn.de}
\bonn\fzj\Tbilisi

\title{Interactions of  the Pseudoscalar Meson Octet and the Baryon Decuplet\\ in the Continuum and a Finite Volume}

\begin{abstract}
This study focuses on the interaction of the pseudoscalar meson octet and the baryon decuplet. In the continuum, it is observed that several $J^{P}=\frac32^-$ baryon resonances can be produced by the Weinberg-Tomozawa interaction in unitarized chiral perturbation theory, including the $N(1875)$, $\Sigma(1670)$, $\Sigma(1910)$, $\Xi(1820)$ and $\Omega(2012)$. Among them, the $\Xi(1820)$ and $\Sigma(1670)$ may exhibit a potential two-pole structures. The unitarized chiral perturbation approach is then applied as the underlying theory to predict the energy levels of these systems in a finite volume. These energy levels are well described by the $K$-matrix parameterization constrained by  flavor SU(3) symmetry. With the parameters from the best fits, the poles extracted from the $K$-matrix parameterization closely correspond to those derived from the underlying chiral effective field theory, as long as they are close to physical region and not significantly higher than the lowest relevant threshold.
\end{abstract}

\maketitle
\section{Introduction}
The interaction between the octet of pseudoscalar mesons (referred to as $\Phi$, which are the Goldstone bosons resulting from the spontaneous breaking of chiral symmetry) and matter fields can be effectively described using chiral perturbation theory (ChPT). Unitary treatments of the chiral dynamics (for early reviews, see~\cite{Oller:2019opk,Oller:2000ma}) can well reproduce the physical resonances as dynamical generated poles, including the $f_0(500)$, $f_0(980)$ and $a_0(980)$ in Ref.~\cite{Oller:1997ti}, the $\Lambda(1405)$ in Refs.\cite{Kaiser:1995eg,Oset:1997it,Oller:2000fj,Jido:2003cb,Garcia-Recio:2002yxy}, the low-lying axial-vector mesons in Ref.~\cite{Roca:2005nm}, the $D_0^*(2300)$ in Refs.~\cite{Kolomeitsev:2003ac,Guo:2006fu,Guo:2009ct} and so on. 

In unitarized ChPT (UChPT) a characteristic finding is that certain systems may have two poles with overlapping signals on the real axis. This phenomenon is commonly referred to as the two-pole structure in literature, with notable examples being the $\Lambda(1405)$ and $D_0^*(2300)$. For a recent review on such two-pole structures, we refer to Ref.~\cite{Meissner:2020khl}. We also note the recent work~\cite{Xie:2023cej} where two-pole structures in the isospin-1/2 $K\Sigma_c$-$\pi \Xi_c^\prime$ coupled channel system where discussed.
Additionally, in Ref.~\cite{Sarkar:2004jh}, it was observed that the $S$-wave interaction between the $\Phi$ and the ground $J^P=\frac32^+$ baryon decuplet (referred to as $\bm T$ hereafter) within the UChPT may result in potential two-pole structures in the $(S,I)=(-1,1)$ and $(-2,\frac12)$ channels, where $S$ and $I$ represent strangeness and isospin, respectively. 

On the other hand, lattice QCD is increasingly important in the study of the hadron spectrum. The energy levels obtained on the lattice are typically fitted using the $K$-matrix parameterization, as shown in Refs.~\cite{Wilson:2014cna,Moir:2016srx}. When dealing with coupled channels, the $K$-matrix parameterization often involves a large number of parameters if no symmetry is taken into account. In Ref.~\cite{Moir:2016srx}, the energy levels of the $D\pi$, $D\eta$, and $D_s\bar K$ coupled channels from the lattice study were fitted, resulting in nine sets of parameters with acceptable $\chi^2/\rm d.o.f.$ in the $S$-wave case. In addition to a reliable bound state below the $D\pi$ threshold, there is another higher pole in the complex plane. However, the precise position of this pole has large uncertainties and is not consistent across different solutions. This issue is addressed in Ref.~\cite{Asokan:2022usm}, where flavor SU(3) symmetry is taken into account when constructing the $K$-matrix. This means that the parameters in the $K$-matrix are not completely independent, leading to a significant reduction in the number of free parameters. As a result, the higher pole becomes more stable, providing support for the two-pole structure of the $D_0^*(2300)$~\cite{Albaladejo:2016lbb}.

In this study, we aim to reevaluate the interaction between $\Phi$ and $\bm T$ using a similar approach as outlined in Ref.~\cite{Sarkar:2004jh}. However, we will make slight adjustments to the subtraction constants in the meson-baryon loop functions. We will then compare the dynamically generated poles with the resonances listed in the Review of Particle Physics (RPP)~\cite{ParticleDataGroup:2022pth}. Additionally, we will investigate the behavior of the $\Phi \bm T$ systems in a finite volume. By considering UChPT as the underlying framework, we can predict the energy levels of the $\Phi \bm T$ systems. Subsequently, we will fit these energy levels using the amplitude parameterized by the $K$-matrix, incorporating constraints from flavor SU(3) symmetry. This analysis will allow us to assess whether the SU(3) constrained $K$-matrix parameterization adequately describes the energy levels in finite volume and accurately reproduces the pole positions. Ultimately, our findings will serve as valuable guidance for future lattice studies on such systems.

\section{Resonances from UChPT}
The interaction between the pseudoscalar meson octet and baryon decuplet can be elucidated using chiral perturbation theory. Specifically, at the lowest order, the Lagrangian can be expressed as follows~\cite{Jenkins:1991es,Jenkins:1991ts,Hemmert:1997ye,Fettes:2000bb} 
\begin{align}
    \mathcal{L}=-i\bar T^{\mu}_{abc}\slashed{\mathcal D}T_{\mu}^{abc}\label{eq:Lag}
\end{align}
where $T^{abc}_{\mu}$ represents the baryon decuplet, with $a,b,c\in\{1,2,3\}$ flavor indices. It is important to note that $T^{abc}_{\mu}$ is fully symmetric in the flavor indices. The identification of $T^{abc}_{\mu}$ with the physical states is\footnote{The negative signs associated with the $\Sigma^*$ and $\Omega$ states can be attributed to the phase convention used in the SU(3) Clebsch–Gordan coefficients in Ref.~\cite{McNamee:1964xq}.}
\begin{align} 
T^{111}&=\Delta^{++}, T^{112}=\frac{ \Delta^{+}}{\sqrt{3}}, T^{122}=\frac{\Delta^0}{\sqrt{3}} , T^{222}=\Delta^{-},\notag\\
T^{113}&=-\frac{\Sigma^{*+}}{\sqrt{3}} , T^{123}=-\frac{ \Sigma^{* 0}}{\sqrt{6}},
 T^{223}=-\frac{\Sigma^{*-}}{\sqrt{3}} , \notag\\
 T^{133}&=\frac{ \Xi^{* 0}}{\sqrt{3}}, T^{233}=\frac{\Xi^{*-}}{\sqrt{3}} , T^{333}=-\Omega^{-} .\label{eq:phcon}
\end{align}
The covariant derivative is expressed as
\begin{align}
    \mathcal{D}^\nu T_{a b c}^\mu=\partial^\nu T_{a b c}^\mu+\left(V^\nu\right)_a^d T_{a b c}^\mu+\left(V^\nu\right)_b^d T_{a d c}^\mu+\left(V^\nu\right)_c^d T_{a b d}^\mu
\end{align}
where the vector current is 
\begin{align}
    V^\mu&=\frac{1}{2}\left(\xi \partial^\mu \xi^{\dagger}+\xi^{\dagger} \partial^\mu \xi\right)
\end{align}
with
\begin{align}
    \xi&=\exp\left({\frac{i\Phi}{\sqrt2 f_\pi}}\right)~,\\
    \Phi&=\left(\begin{array}{ccc}
\frac{1}{\sqrt{2}} \pi^0+\frac{1}{\sqrt{6}} \eta & \pi^{+} & K^{+} \\
\pi^{-} & -\frac{1}{\sqrt{2}} \pi^0+\frac{1}{\sqrt{6}} \eta & K^0 \\
K^{-} & \bar{K}^0 & -\frac{2}{\sqrt{6}} \eta
\end{array}\right)
\end{align}
and $f_\pi=92.4$~MeV the pion decay constant.

The potential of the interaction between $\Phi$ and $T$ can be determined by utilizing the Lagrangian equation provided in Eq.\eqref{eq:Lag}. The transition from channel $i$ to channel $j$ can be represented as stated in the following reference~\cite{Sarkar:2004jh,Oset:1997it}\footnote{The convention used for $V_{ij}$ differs from that in Ref.~\cite{Sarkar:2004jh} by a factor of $\sqrt{4m_im_j}$. Consequently, the loop function in Eq.~\eqref{eq:GDR} is different from that in Ref.~\cite{Sarkar:2004jh} by a factor of $2m$.},
\begin{align}
    V_{ij}=-\frac{C_{ij}}{4f_\pi^2}\sqrt{4m_im_j}(k^0+k^{\prime 0})\label{eq:Vij},
\end{align}
where $k$ and $k^{\prime}$ denote the momenta of the incoming and outgoing mesons, respectively. The constant $C_{ij}$ relies on the specific incoming and outgoing channels, and can be determined by utilizing the Lagrangian equation provided in Eq.~\eqref{eq:Lag}\footnote{Note that several $C_{ij}$ are opposite to those in Ref.~\cite{Sarkar:2004jh} due to the phase conventions in Eq.~\eqref{eq:phcon}.}. As mostly done in the literature, we neglect here the Born terms, that are formally also of leading order. {We refer to Refs.~\cite{Bruns:2010sv,Mai:2020ltx} for some detailed discussions.}

By utilizing the potential defined in Eq.~\eqref{eq:Vij}, we can solve the on-shell Bethe-Salpeter equation
\begin{align}
    T=(1-VG)^{-1}V
\end{align}
to search for poles of the $S$-wave $\Phi \bm T$ system. The two point loop function for a meson with mass $M$ and a baryon with mass $m$ reads~\cite{Oller:2000fj}
\begin{align}
    G(s)&=i \int \frac{d^4 q}{(2 \pi)^4} \frac{1}{(P-q)^2-m^2+i \epsilon} \frac{1}{q^2-M^2+i \epsilon}\notag\\
    =&\,\frac{2m}{16\pi^2}\bigg\{a(\mu)+\log\frac{m^2}{\mu^2}+\frac{s-\Delta}{2s} \log\frac{M^2}{m^2} \nonumber\\
    &+\frac{k}{\sqrt s} \Big[\log\left(2k\sqrt s+s+\Delta\right) + \log\left(2k\sqrt s+s-\Delta\right)  \notag\\
    &-\log\left(2k \sqrt s-s+\Delta\right) - \log\left(2k \sqrt s-s-\Delta\right)\Big]\bigg\},\label{eq:GDR}
\end{align}
where $\Delta=m^2-M^2$, $k$ represents the 3-momentum of the meson in the center of mass frame, and $a(\mu)$ is a subtraction constant with $\mu$ the scale of dimensional regularization (DR). In Ref.~\cite{Sarkar:2004jh}, the subtraction constant is assigned a value of $-2$ for $\mu=700$~MeV. Subsequently, several poles of the scattering amplitude $T$ are found, some of which may provide an explanation for the $J^P=\frac{3}{2}^-$ baryons in the RPP~\cite{ParticleDataGroup:2022pth}. In this paper, we follow the approach outlined in Refs.~\cite{Lutz:2001yb,Hyodo:2008xr,Guo:2023wes}, where the loop function satisfies 
\begin{align}
    G(s=m^2;a(\mu))=0,
\end{align}
and accordingly, $a(\mu)$ is determined for a given $\mu$. It is worth noting that this approach is consistent with Ref.~\cite{Sarkar:2004jh} in the SU(3) symmetry case, albeit slightly different for the physical masses.

In order to search for poles of the scattering amplitude, it is necessary to examine the complex energy plane. The presence of intermediate states that satisfy the on-shell condition leads to what is known as right-hand cuts. By crossing the cut to the unphysical Riemann sheet (RS) of a specific channel, the loop function undergoes the following modification,
\begin{align}
    G^{(-)}(s)=G^{(+)}(s)+2im\frac{k}{4\pi\sqrt s}\label{eq:G2},
\end{align}
where $G^{(+)}(s)$ represents the loop function on the physical RS defined in Eq.\eqref{eq:GDR} and $k$ denotes the 3-momentum in the center-of-mass frame. In the case of $n$ coupled channels, there are a total of $2^n$ RSs, which can be distinguished by the labels $(\pm,\pm,\cdots,\pm)$, where the $i$-th $\pm$ indicates the loop function of the $i$-th channel. For a nice pictorial of this, see e.g. Fig.~3 in Ref.~\cite{Mai:2022eur}.

Based on the flavor SU(3) symmetry, we can decompose the $\Phi \bm T$ system in flavor space as follows,
\begin{align}
    10\otimes8=8\oplus10\oplus27\oplus35.
\end{align}
The corresponding coefficients in the potential are
\begin{align}
    C_{ij}={\rm diag}(6,3,1,-1),
\end{align}
as obtained in Refs.\cite{Sarkar:2004jh,Kolomeitsev:2003kt}. These suggest that the octet and decuplet have strong attractions, the 27-plet exhibits a weak attraction while the 35-plet is repulsive. Therefore, we expect the presence of near-threshold poles in the octet and decuplet systems. After the breaking of SU(3) symmetry, channels in different multiplets with the same strangeness and isospin $(S,I)$ can couple with each other. In the following discussion, we will focus on the six combinations of $(S,I)$ that appear in the octet and decuplet systems. The possible poles in each $(S,I)$ sector are listed in Table~\ref{tab:poles}, along with the possible corresponding resonances collected in the RPP~\cite{ParticleDataGroup:2022pth}. In the following, we provide some detailed discussions for each sector.
\begin{table*}[t]
\caption{The poles of $\Phi \bm T$ systems prediced by UChPT and the possibly corresponding $J^P=\frac32^-$ baryons as listed in the RPP~\cite{ParticleDataGroup:2022pth}.}\label{tab:poles}
\begin{tabular}{|c|c|cc|cc|}
\hline\hline
\multirow{2}{*}{$(S,I)$}        & \multirow{2}{*}{Threshold (MeV)}            & \multicolumn{2}{c|}{UChPT}                   & \multicolumn{2}{c|}{RPP}                           \\ \cline{3-6} 
                                &                                             & \multicolumn{1}{c|}{Pole (MeV)}        & RS       & \multicolumn{1}{c|}{\ Resonance\ }      & Pole (MeV)       \\ \hline
\multirow{2}{*}{$(0,\frac12)$}  & \multirow{2}{*}{$(1370, 1880)$}             & \multicolumn{1}{c|}{$1387-77i$}  & $(-+)$   & \multicolumn{1}{c|}{}      &   \\
                                &                                             & \multicolumn{1}{c|}{$1883+12i$}  & $(+-)$   & \multicolumn{1}{c|}{$N(1875)$}      & $\ \ 1900-80i\ \ $  \\ \hline
\multirow{3}{*}{$(0,\frac32)$}  & \multirow{3}{*}{$(1370, 1780, 1880)$}       & \multicolumn{1}{c|}{$1477-194i$} & $(-++)$  & \multicolumn{1}{c|}{}               &             \\
                                &                                             & \multicolumn{1}{c|}{$1909-19i$}  & $(--+)$  & \multicolumn{1}{c|}{}               &             \\
                                &                                             & \multicolumn{1}{c|}{$1527-247i$} & $(---)$  & \multicolumn{1}{c|}{}               &             \\ \hline
\multirow{2}{*}{$(-1,0)$}       & \multirow{2}{*}{$(1523, 2029)$}             & \multicolumn{1}{c|}{$1560-97i$}  & $(-+)$   & \multicolumn{1}{c|}{}               &             \\
                                &                                             & \multicolumn{1}{c|}{$2033-18i$}  & $(-+)$   & \multicolumn{1}{c|}{}               &             \\ \hline
\multirow{4}{*}{$\ \ (-1,1)\ \ $}       & \multirow{4}{*}{$\ (1523, 1728, 1932, 2029)\ $} & \multicolumn{1}{c|}{$1609-8i$}   & $(-+++)$ & \multicolumn{1}{c|}{$\Sigma(1670)$} & $1665-28i$  \\
                                &                                             & \multicolumn{1}{c|}{$\ \ 1664-194i\ \ $} & $\ \ (-+++)\ \ $ & \multicolumn{1}{c|}{}               &             \\
                                &                                             & \multicolumn{1}{c|}{$1986-11i$}  & $(--++)$ & \multicolumn{1}{c|}{}               &             \\
                                &                                             & \multicolumn{1}{c|}{$1965-62i$}  & $(---+)$ & \multicolumn{1}{c|}{$\Sigma(1910)$} & $1910-110i$ \\ \hline
\multirow{3}{*}{$(-2,\frac12)$} & \multirow{3}{*}{$(1671, 1880, 2081, 2168)$} & \multicolumn{1}{c|}{$1806-16i$}  & $(-+++)$ & \multicolumn{1}{c|}{$\Xi(1820)$}    & $1823-12i$  \\
                                &                                             & \multicolumn{1}{c|}{$1818-170i$} & $(-+++)$ & \multicolumn{1}{c|}{}               &             \\
                                &                                             & \multicolumn{1}{c|}{$2093-51i$}  & $(---+)$ & \multicolumn{1}{c|}{}               &             \\ \hline
$(-3,0)$                        & $(2029, 2220)$                              & \multicolumn{1}{c|}{$2016$}      & $(++)$   & \multicolumn{1}{c|}{$\Omega(2012)$} & $2012-3i$   \\ \hline\hline
\end{tabular}
\end{table*}

For the case of $(S,I)=(0,1/2)$, there exist two channels, namely ($\pi\Delta,\ K\Sigma^*$), with thresholds (1370, 1880)~MeV. Utilizing the coefficient matrix $C_{ij}$ where
\begin{align}
    C_{ij}=\left(\begin{array}{cc}
        5 & -2 \\
         -2 & 2
    \end{array}\right),
\end{align}
one can find two poles. One pole is located at $1387-77i$~MeV on RS $(-+)$, directly linked to the physical region and resulting in a broad peak on the physical axis. In the RPP, the closest baryon to this pole is the $N(1520)$, situated approximately at $1510-55i$~MeV. We do not propose to match these states here as done in~\cite{Sarkar:2004jh}, as a better UChPT description based on Goldstone boson scattering of the baryon ground state octet can be obtained~\cite{Bruns:2010sv}. 
The second pole, found at $1883+12i$~MeV on RS $(+-)$, is not directly connected to the physical region. Its influence on $|T_{K\Sigma^*\to K\Sigma^*}|$ is reflected as a peak-like threshold cusp exactly at the $K\Sigma^*$ threshold~\cite{Dong:2020hxe}. In the RPP, there is a nearby baryon $N(1875)$ located around $1900-80i$~MeV. The threshold cusp in our model provides a possible explanation for the experimental signal of the $N(1875)$. It is found in Refs.~\cite{He:2012ud,He:2014gga} that there is no place for $N(1875)$ being a three quark state in the constituent quark model.  A detailed study of $N(1875)$ from the $K\Sigma^*$ interaction, together with several coupled channels and possible triangle singularity, is presented in Ref.~\cite{Samart:2017scf}. In Refs.~\cite{He:2015yva,He:2017aps}, the $N(1875)$ is identified as a $K\Sigma^*$ bound state in the one-boson-exchange model. A recent analysis~\cite{Ronchen:2022hqk} on the pion- and photon-induced production of $\pi N,\eta N,K\Lambda$ and $K\Sigma$ final states within the J\"ulich-Bonn model also indicates a dynamically generated state in $D_{13}$ partial wave, whose mass is close to $N(1875)$ but the total width ($\Gamma>600$ MeV) is much larger.

For the case of $(S,I)=(0,3/2)$, there are three channels, namely $\pi\Delta$, $\eta\Delta$ and $K\Sigma^*$, with corresponding thresholds at 1370, 1780, and 1880~MeV. Utilizing the matrix $C_{ij}$ where
\begin{align}
    C_{ij}=\left(\begin{array}{ccc}
        2 & 0 & -\sqrt{5/2} \\
        0 & 0 & -3/\sqrt2 \\
        -\sqrt{5/2} & -3/\sqrt2 & -1
    \end{array}\right),
\end{align}
one can find three poles, located at $1477-194i$~MeV on the RS $(-++)$, $1909-19i$~MeV on the RS $(--+)$ and $1527-247i$~MeV on the RS $(---)$. The first two poles can leave imprints on the scattering amplitude whereas the third pole lies far outside the physical region. However, there are no suitable candidates in the RPP for the first two poles.

For the case of $(S,I)=(-1,0)$, there are two channels, namely, $\pi\Sigma^*$ and $K\Xi^*$, with corresponding thresholds of 1523 and 2029~MeV. By utilizing the matrix $C_{ij}$ given by
\begin{align}
    C_{ij}=\left(\begin{array}{cc}
        4 & \sqrt6 \\
         \sqrt6 & 3
    \end{array}\right),
\end{align}
one can find two poles, both located on the RS $(-+)$, one at $1560-97i$~MeV and the other at $2033-18i$~MeV. However, there is no state in the RPP that corresponds to these two poles.

For the case of $(S,I)=(-1,1)$, there are four channels, namely, $\pi\Sigma^*$, $\bar K\Delta$, $\eta\Sigma^*$ and $K\Xi^*$, with with respective thresholds at 1523, 1728, 1932 and 2029~MeV. By utilizing the matrix $C_{ij}$ given by
\begin{align}
    C_{ij}=\left(\begin{array}{cccc}
        2 & 1 & 0 & -2 \\
        1 & 4 & -\sqrt6 & 0 \\
        0 & -\sqrt6 & 0 & -\sqrt6 \\
        2 & 0 & -\sqrt6 & 1 
    \end{array}\right),
\end{align}
one can find four poles. The one located at $1609-8i$~MeV on the RS $(-+++)$ may correspond to the $\Sigma(1670)$ at $1665-30i$~MeV in the RPP. On the same RS, there exists a broader pole at $1664-194i$~MeV, which has the potential to distort the narrower pole and may result in a two-pole structure. The pole at $1986-11i$~MeV on the RS $(--++)$ is not directly linked to the physical axis and therefore has a negligible effect. Conversely, the poles at $1965-62i$~MeV on the RS of $(---+)$ are directly connected to the physical axis and can be matched to the $\Sigma(1910)$ at $1910-110i$~MeV in RPP.

For the case of $(S,I)=(-2,1/2)$, there are four channels, namely $\pi\Xi^*$, $\bar K\Sigma^*$, $\eta\Xi^*$ and $K\Omega$, respective thresholds at 1671, 1880, 2081, and 
2168~MeV. By utilizing the matrix $C_{ij}$ given by
\begin{align}
    C_{ij}=\left(\begin{array}{cccc}
        2 & 1 & 0 & 3/\sqrt2 \\
        1 & 2 & -3 & 0 \\
        0 & -3 & 0 & -3/\sqrt2 \\
        3/\sqrt2 & 0 & -3/\sqrt2 & 3 
    \end{array}\right),
\end{align}
one can find three poles. The pole at $1806-16i$~MeV on the RS $(-+++)$ is close proximity to the $\Xi(1820)$ in the RPP. Similarly to the $(S,I)=(-1,1)$ case, there is another broader pole at $1818-170i$~MeV on the same RS, which has a similar mass but significantly different width. This suggests the presence of a possible two-pole structure. It has been pointed out in Ref.~\cite{Molina:2023uko} that these two poles can effectively describe the experimental data in the vicinity of the $\Xi(1820)$ region~\cite{BESIII:2023mlv}. The pole at $2093-51i$~MeV on RS $(---+)$ can lead to observable peaks around 2100~MeV, although as of now, no corresponding state has been identified in the RPP.

For the case of $(S,I)=(-3,0)$, there are two channels, namely $\bar K\Xi^*$ and $\eta\Omega$, with corresponding thresholds at 2029 and 2220~MeV. By utilizing the matrix $C_{ij}$ given by
\begin{align}
    C_{ij}=\left(\begin{array}{cc}
        0 & 3 \\
         3 & 0
    \end{array}\right),
\end{align}
one can find a bound state at $2016$~MeV on RS $(++)$.  This finding is consistent with the recently observed $\Omega(2012)$ reported in the Belle experiment~\cite{Belle:2018mqs}, as also discussed in Refs.~\cite{Valderrama:2018bmv,Pavao:2018xub,Lu:2020ste,Ikeno:2020vqv}

Please note that the poles we have obtained are qualitatively consistent with those in Ref.~\cite{Sarkar:2004jh}, although the precise pole positions differ. It is important to emphasize that the positions of the poles depend significantly on the subtraction constants $a(\mu)$, and for each channel, $a(\mu)$ may need to be adjusted slightly in order to match the observed resonances. The dynamically generated poles from UChPT that are still missing can be searched for in future experiments and lattice studies.

\section{Energy levels in finite volume from UChPT}
In order to examine the energy levels of the $\Phi \bm T$ systems within a finite volume, we have to modify the loop function used in the previous section. Instead of using dimensional regularization, we introduce a form factor to regularize the loop function, which is given by:
\begin{align}
    &G(s)=i \int \frac{d^4 q}{(2 \pi)^4} \frac{1}{(P-q)^2-m^2+i \epsilon} \frac{1}{q^2-M^2+i \epsilon}\notag\\
    &\to\int \frac{d^3 \bm q}{(2 \pi)^3} \frac{ F(|\bm q|) }{2\omega_m(\bm q)\omega_M(\bm q)} \frac{\omega_m(\bm q)+\omega_M(\bm q)}{s-(\omega_m(\bm q)+\omega_M(\bm q))^2+i \epsilon}\label{eq:Gff}
\end{align}
where $\omega_m(\bm q)=\sqrt{\bm q^2+m^2}$, $\omega_M(\bm q)=\sqrt{\bm q^2+M^2}$ and $P=(\sqrt s,\bm 0)$. The form factor, as chosen in Refs.~\cite{Doring:2011vk,Asokan:2022usm}, is given by
\begin{align}
    F(q)=\frac{\Lambda^{12}}{\Lambda^{12}+q^{12}}
\end{align}
with the cutoff parameter $\Lambda=1.2$~GeV. Unitarity dictates that $F(q)$ must be equal to 1 when the momentum $q$ is on-shell. To ensure that the difference $F(q)-1$ is less than 1\% within the energy range of interest in the subsequent analysis, the power $12$ is chosen.
It is worth noting that with this form factor, the loop function is significantly larger than the one used in last section, which in turn leads to the presence of bound states in certain channels. In lattice studies, the pion mass is typically larger than its physical mass. In ChPT, the interaction between Goldstone bosons and matter fileds becomes stronger as the masses of the Goldstone bosons increase. Therefore, in the following, we will retain the physical masses of the pseudoscalar mesons and utilize the larger loop function given in Eq.~\eqref{eq:Gff}.

\begin{figure}[th]
    \centering
    \includegraphics[width=\linewidth]{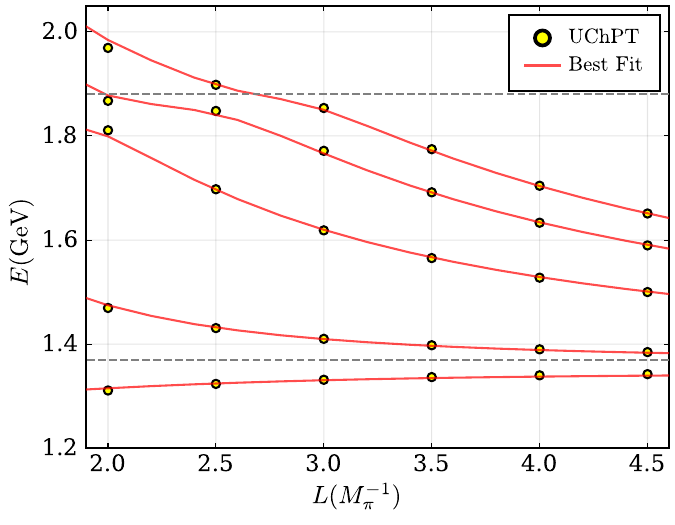}
    \caption{The lowest energy levels in a cubic volume of finite size of length $L$, for the systems with $(S,I)=(0,1/2)$. The yellow circles represent the predictions incorporating interaction from the UChPT. The red line represents the best fit achieved using the SU(3) constrained $K$-matrix parametrization. The gray dashed lines denotes the relevant thresholds in close proximity.}
    \label{fig:S0Ip5}
\end{figure}

\begin{figure}[th]
    \centering
    \includegraphics[width=\linewidth]{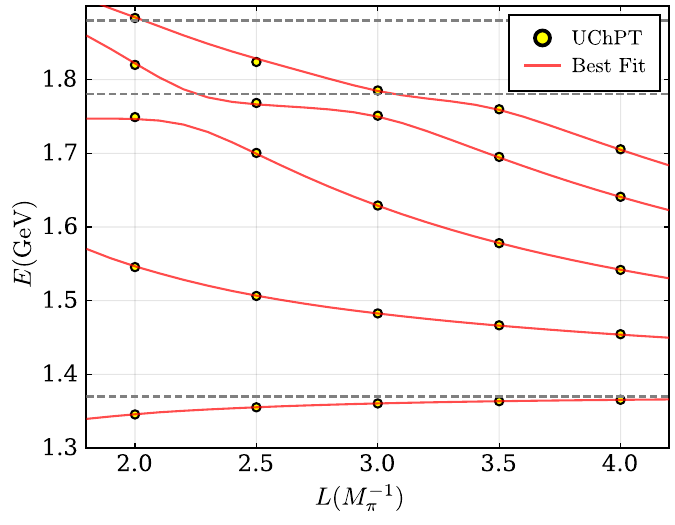}
    \caption{The lowest energy levels in a cubic volume of finite size with a length of $L$ for systems with $(S,I)=(0,3/2)$. See the caption of Fig.~\ref{fig:S0Ip5}. }
    \label{fig:S0I1p5}
\end{figure}

\begin{figure}[th]
    \centering
    \includegraphics[width=\linewidth]{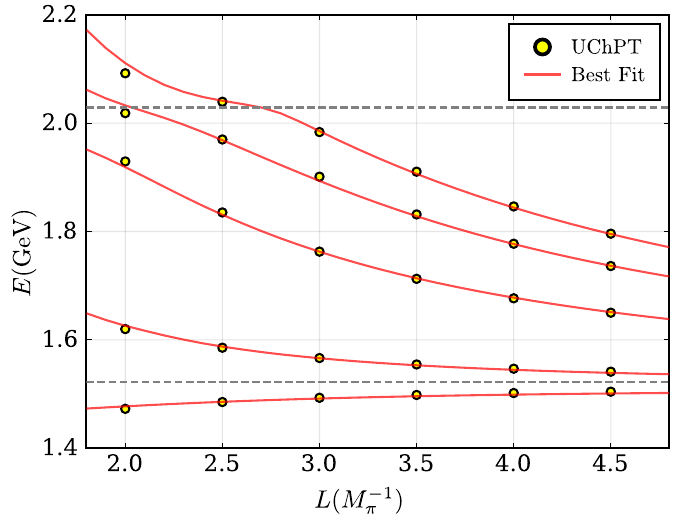}
    \caption{The lowest energy levels for the systems with $(S,I)=(-1,0)$. See the caption of Fig.~\ref{fig:S0Ip5}. }
    \label{fig:Sm1I0}
\end{figure}

\begin{figure}[th]
    \centering
    \includegraphics[width=\linewidth]{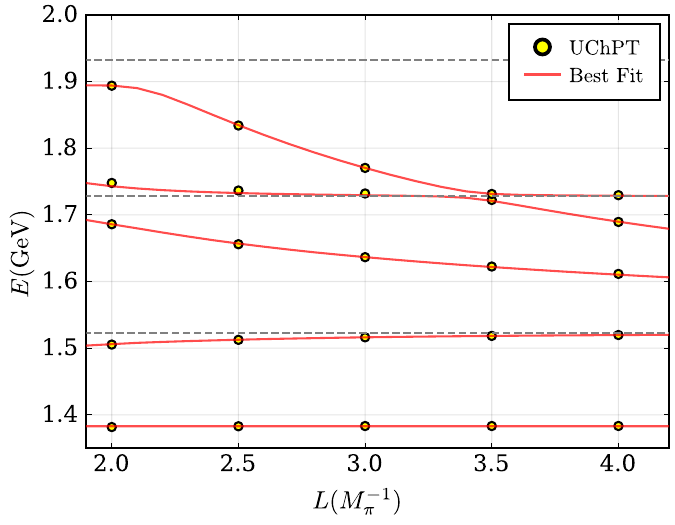}
    \caption{The lowest energy levels for the systems with $(S,I)=(-1,1)$. See the caption of Fig.~\ref{fig:S0Ip5}. }
    \label{fig:Sm1I1}
\end{figure}

\begin{figure}[th]
    \centering
    \includegraphics[width=\linewidth]{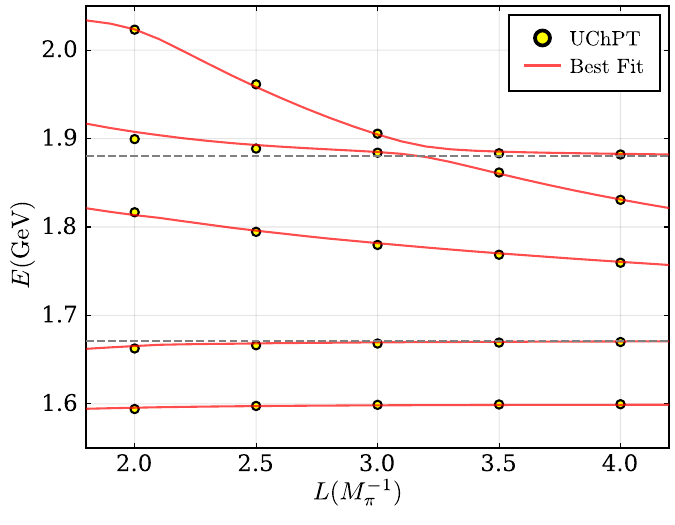}
    \caption{The lowest energy levels for the systems with $(S,I)=(-2,1/2)$. See the caption of Fig.~\ref{fig:S0Ip5}. }
    \label{fig:m2I0p5}
\end{figure}

\begin{figure}[th]
    \centering
    \includegraphics[width=\linewidth]{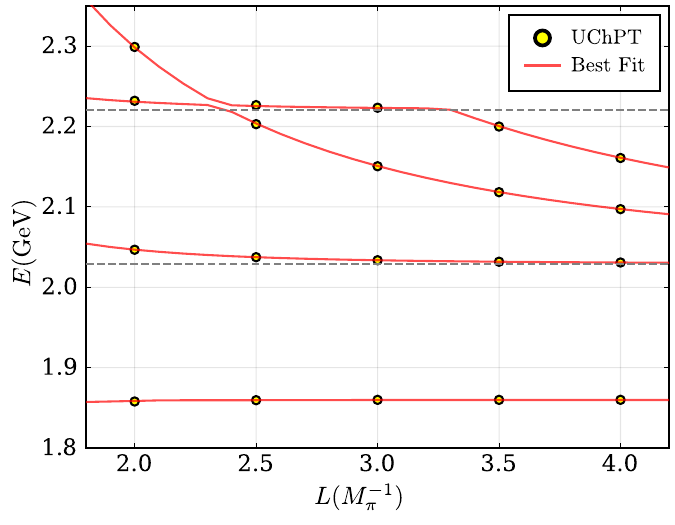}
    \caption{The lowest energy levels for the systems with $(S,I)=(-3,0)$. See the caption of Fig.~\ref{fig:S0Ip5}. }
    \label{fig:Sm3I0}
\end{figure}

The energy levels in finite volume are determined by the poles of the scattering amplitude $\tilde T$, which can be solved using the equation
\begin{align}
    \tilde T=(1-\tilde V\tilde G)^{-1}\tilde V\label{eq:Ttilde}
\end{align}
where the loop function in finite volume reads
\begin{align}
    \tilde G=\frac{1}{L^3}\sum_{\bm q} \frac{F(|\bm q|)}{2\omega_m(\bm q)\omega_M(\bm q)} \frac{\omega_m(\bm q)+\omega_M(\bm q)}{s-(\omega_m(\bm q)+\omega_M(\bm q))^2}\label{eq:Gsum}
\end{align}
with $L$ the length of the spatial space and 
\begin{align}
    \bm q=\frac{2\pi}{L}\bm n,\ \bm n\in \mathbb Z^3.
\end{align}
The potential in the finite volume, denoted as $\tilde V$, is the same as that in the continuous case, with exponentially suppressed corrections~\cite{Luscher:1986pf,Gasser:1987zq}. Therefore, we can use Eq.\eqref{eq:Vij} for $\tilde V$. The lowest energy levels for each $(S,I)$ sector are represented by the yellow circles in Figs.~(\ref{fig:S0Ip5}-\ref{fig:Sm3I0}).


\section{Fitting with a SU(3) $K$-matrix parameterization}
The energy levels obtained from lattice calculations in finite volume are often analyzed by fitting them with a parameterized scattering amplitude using the $K$-matrix approach. For example, in Refs.~\cite{Wilson:2014cna,Moir:2016srx}, the $S$-wave coupled-channel scattering amplitudes are parameterized as
\begin{align}
    t_{ij}^{-1}(s)=K_{ij}^{-1}(s)+I_{ij}^{}(s)\label{eq:tijinv}
\end{align}
where 
\begin{align}
    K_{ij}(s)&=\frac{\left(g_i^{(0)}+g_i^{(1)}s\right)\left(g_j^{(0)}+g_j^{(1)}s\right)}{m^2-s}+\gamma_{ij}^{(0)}+\gamma_{ij}^{(1)}s,\label{eq:K}\\
    I_{ij}(s)&=\delta_{ij}\left(I^{(i)}_{\rm CM}(s)-I^{(i)}_{\rm CM}(m^2)\right).\label{eq:Iij}
\end{align}
Here, $m$, $g_{i}^{(0,1)}$ and $\gamma_{ij}^{(0,1)}$ are parameters. The imaginary part of $t_{ij}^{-1}(s)$ is constrained by the unitarity and $I_{\rm CM}(s)$ is the Chew-Mandelstam prescription~\cite{Chew:1960iv},
\begin{align}
    I^{(i)}_{\rm CM}(s)&=\frac{\rho_i(s)}{\pi}\log\left[\frac{\xi_i(s)+\rho_i(s)}{\xi_i(s)-\rho_i(s)}\right]\notag\\
    &-\frac{\xi_i(s)}{\pi}\frac{m_2^{(i)}-m_1^{(i)}}{m_2^{(i)}+m_1^{(i)}}\log\frac{m_2^{(i)}}{m_1^{(i)}}
\end{align}
with
\begin{align}
& \xi_i(s)=1-{\left(m_1^{(i)}+m_2^{(i)}\right)^2}/{s}, \\
& \rho_i^2(s)=\xi_i(s)\left(1-{\left(m_1^{(i)}-m_2^{(i)}\right)^2}/{s}\right),
\end{align}
where $m_1^{(i)}$ and $m_2^{(i)}$ are the masses of the two particles in channel $i$. Note that $I_{ij}(s)$ in Eq.~\eqref{eq:Iij} are subtracted at the mass of the bound state, if it exists. Otherwise, if no bound state is present, the subtraction points will be chosen at the corresponding threshold.

From Eq.~\eqref{eq:Ttilde}, the scattering amplitude with the $K$-matrix parameterization in finite volume can be expressed as, 
\begin{align}
    \tilde T_K&=\frac{1}{\tilde V^{-1}-\tilde G}=\frac1{T_K^{-1}+G-\tilde G}\notag\\
    &=\frac{1}{-t^{-1}/(16\pi)+G-\tilde G},\label{eq:TKtilde}
\end{align}
where $t^{-1}=-16\pi T_K^{-1}$ is defined in Eq.~\eqref{eq:tijinv}. After the parameters are determined by the fit procedure, we can extract the pole positions of $\tilde T_K$.

In the $K$-matrix parameterization, the RS is determined by the imaginary part of $\rho_i(s)$. More precisely, in the case of $n$ coupled channels, the scattering amplitude's $2^n$ RSs are identified by $(\pm,\pm,\cdots,\pm)$, where the $i$-th $\pm$ stands for the sign of the imaginary part of $\rho_i(s)$. Such a convention is consistent with Eq.~\eqref{eq:G2}.

\subsection{Flavor SU(3) symmetry}
The $K$-matrix parameterization mentioned above contains a large number of parameters. In order to reduce the number of free parameters, we adopt the approach outlined in Ref.~\cite{Asokan:2022usm} by modifying the $K$-matrix in Eq.~\eqref{eq:K} as follows:
\begin{align}
K^{(S,I)}=\sum_{R}\left(\frac{\left(g^{(S,I)}_{R}\right)^2}{\left(m^{(S,I)}_{R}\right)^2-s}+c^{(S,I)}_R\right)C_R^{(S,I)},\label{eq:KR}
\end{align}
where $g_R^{(S,I)}$, $c_R^{(S,I)}$ and $m_R^{(S,I)}$ are real parameters and $C_R$ is a constant matrix that depends on the representation of SU(3) and the $(S,I)$ of the system. For a given $(S,I)$, the coupled channels can be decomposed into the direct sum of several multiplets of the SU(3) group, and $R$ in Eq.~\eqref{eq:KR} represents all multiplets involved in the decomposition. We have observed that for $R=8$ and $10$, there exist near threshold poles, while for $R=27$ and $35$, the interaction is too weak to generate poles. Therefore, we will not include the pole term in Eq.~\eqref{eq:KR}, specifically setting $g^{(S,I)}_{27}=g^{(S,I)}_{35}=0$. 

For the case of $(S,I)=(0,1/2)$, the relation between the isospin-symmetric particle basis ($\ket{\pi\Delta},\ \ket{K\Sigma^*}$) and the SU(3) flavor basis ($\ket{8},\ \ket{27}$) is given by
\begin{align}
    \left(\begin{array}{c}
|8\rangle \\
|27\rangle 
\end{array}\right)=U\left(\begin{array}{c}
|\pi\Delta\rangle \\
|K\Sigma^*\rangle
\end{array}\right),
\end{align}
where the transformation matrix $U$ can be read from the SU(3) CG coefficients~\cite{McNamee:1964xq},
\begin{align}
    U=\sqrt{\frac15}\left(
\begin{array}{cccc}
 -2 & 1 \\
 1 & 2 \\
\end{array}
\right).
\end{align}
By applying this transformation, we can obtain the following expressions
\begin{align}
        C_{8}^{(0,1/2)}&=\frac15 \left(
\begin{array}{cc}
 4 & -2 \\
 -2 & 1 \\
\end{array}
\right), \\
        C_{27}^{(0,1/2)}&=\frac15 \left(
\begin{array}{cc}
 1 & 2 \\
 2 & 4 \\
\end{array}
\right).
\end{align}
The same procedure can be applied to calculate the coupling structures for the other systems, and the results are summarized in Appendix~\ref{app:CR}.

\subsection{Fit results}
By employing the loop function in equation Eq.~\eqref{eq:Gff}, we can determine the poles of the scattering amplitude for each $(S,I)$ sector. These poles are documented in Table~\ref{tab:polesGFF}. Upon comparison with the poles listed in Table~\ref{tab:poles}, it becomes evident that the employed form factor generates significantly stronger interactions in comparison to the DR form in Eq.~\eqref{eq:GDR}.

The SU(3) constrained $K$-matrix parameterization in Eq.~\eqref{eq:KR} has been used to fit the energy levels of $\Phi \bm T$ systems in finite volumes generated by the UChPT interaction, as shown by the yellow circle in Fig.~(\ref{fig:S0Ip5}-\ref{fig:Sm3I0}). The red solid lines in the same plots represent the obtained best fits. It is evident that that the lowest energy levels in each $(S,I)$ sector can be well described by the $K$-matrix in Eq.~\eqref{eq:KR}. The parameters obtained from the fitting process are given in Appendix~\ref{app:fitpars}.

The poles obtained from Eq.~\eqref{eq:TKtilde} using the best fitted parameters are presented in Table~\ref{tab:polesGFF}. By comparing the poles listed in the third column to those in the fifth column, we can observe that in each $(S,I)$ sector, the poles located near the physical region and with lower masses exhibit better agreement with each other, while the higher the pole is, the greater the difference between the two cases. This phenomenon can be attributed to the different energy dependence exhibited by the scattering amplitude. The $K$-matrix effectively describes the UChPT interaction within a finite energy range, and as the energy range expands, the discrepancy becomes more noticeable.

\begin{figure*}[th]
    \centering
    \includegraphics[width=\linewidth]{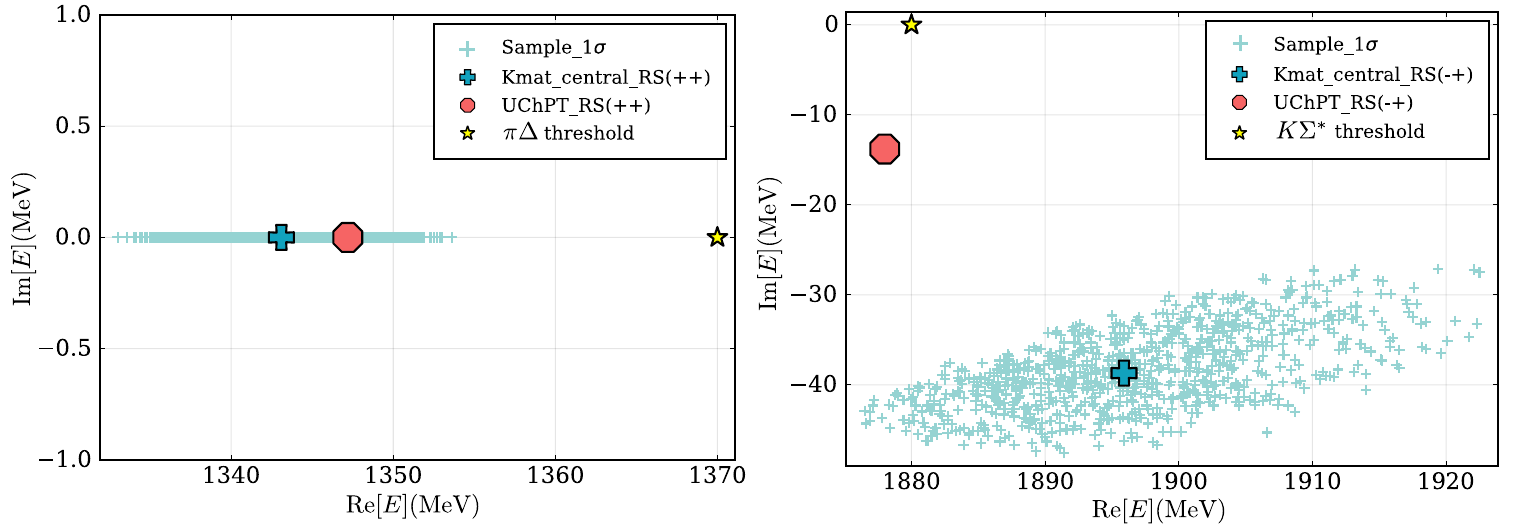}
    \caption{The distribution of the pole positions on RS $(++)$ and $(-+)$ for the systems with $(S,I)=(0,1/2)$ within 1 $\sigma$. }
    \label{fig:pole1rs}
\end{figure*}

To verify the consistency of the poles in the two scenarios, taking into account the uncertainties of the fit procedure, we conducted an error analysis using the $(S,I)=(0,1/2)$ sector as a case study. Since the energy levels were calculated using UChPT interaction without any errors, comparing the $\chi^2/\rm d.o.f.$ directly with 1 and interpreting the uncertainties of the fitted parameters would be inappropriate. It is worth mentioning that the energy level errors in Ref.~\cite{Moir:2016srx} are on the order of 10~MeV, which is typically larger for baryons in lattice studies. Therefore, we assigned an error of 20~MeV to the energy levels of the $\Phi \bm T$ systems. For the particular case of the $(S,I)=(0,1/2)$ sector, the two poles were found to be located at $1343\pm11$~MeV on the RS (++) and at $1896^{+26}_{-19}-14^{+13}_{-9}i$~MeV on the RS $(-+)$. The first pole in both cases exhibits agreement within the uncertainty, while the higher pole does not, as shown in FIG.~\ref{fig:pole1rs}. We have used MINUIT~\cite{James:1975dr,iminuit,iminuit.jl} to fit the data and obtained the errors of the free parameters. According to the correlation matrix and errors, we can produce a number of samples. Then those sets of parameters within $1 \sigma$ defined by theMahalanobis distance~\cite{mdis} are used to calculate the  pole positions. 

\begin{table*}[t]
\caption{The poles of $\Phi \bm T$ systems predicted by the UChPT with loop function in Eq.~\eqref{eq:Gff} and those extracted from the $K$-matrix parameterization by fitting the energy levels in finite volume generated by the UChPT.}\label{tab:polesGFF}
\begin{tabular}{|c|c|cc|cc|}
\hline\hline
\multirow{2}{*}{$(S,I)$}        & \multirow{2}{*}{Threshold (MeV)}            & \multicolumn{2}{c|}{UChPT}                   & \multicolumn{2}{c|}{$K$-matrix}                           \\ \cline{3-6} 
                                &                                             & \multicolumn{1}{c|}{Pole (MeV)}        & RS       & \multicolumn{1}{c|}  {Pole (MeV)}    &    {\ RS\ }    \\ \hline
\multirow{2}{*}{$(0,\frac12)$}  & \multirow{2}{*}{$(1370, 1880)$}             & \multicolumn{1}{c|}{$1347$}  & $(++)$   & \multicolumn{1}{c|}{$1343$}      & $(++)$  \\
                                &                                             & \multicolumn{1}{c|}{$1878-14i$}  & $(-+)$   & \multicolumn{1}{c|}{$1896-39i$}      &  $(-+)$  \\ \hline
\multirow{3}{*}{$(0,\frac32)$}  & \multirow{3}{*}{$(1370, 1780, 1880)$}       & \multicolumn{1}{c|}{$1431-53i$} & $(-++)$  & \multicolumn{1}{c|}{$1427-53i$}               &     $(-++)$        \\
                                &                                             & \multicolumn{1}{c|}{$1784-18i$}  & $(-++)$  & \multicolumn{1}{c|}{$1784-16i$}               &    $(-++)$         \\
                                &                                             & \multicolumn{1}{c|}{$1572-99i$} & $(--+)$  & \multicolumn{1}{c|}{$1589-9i$}               &      $(--+)$       \\ \hline
\multirow{2}{*}{$(-1,0)$}       & \multirow{2}{*}{$(1523, 2029)$}             & \multicolumn{1}{c|}{$1512$}  & $(++)$   & \multicolumn{1}{c|}{$1507$}               &    $(++)$         \\
                                &                                             & \multicolumn{1}{c|}{$2000-48i$}  & $(-+)$   & \multicolumn{1}{c|}{$2004-133i$}               &   $(-+)$          \\ \hline
\multirow{4}{*}{$\ \ (-1,1)\ \ $}       & \multirow{4}{*}{$\ (1523, 1728, 1932, 2029)\ $} & \multicolumn{1}{c|}{$1383$}   & $(++++)$ & \multicolumn{1}{c|}{$1383$} & $(++++)$  \\
                                &                                             & \multicolumn{1}{c|}{$1603-48i$} & $\ \ (-+++)\ \ $ & \multicolumn{1}{c|}{$1596-52i$}               &   $\ \ (-+++)\ \ $          \\
                                &                                             & \multicolumn{1}{c|}{$1925-39i$}  & $(-+++)$ & \multicolumn{1}{c|}{$1922-11i$}               &  $(-+++)$           \\
                                &                                             & \multicolumn{1}{c|}{$1591-46i$}  & $(--++)$ & \multicolumn{1}{c|}{$1603-46i$} & $(--++)$
                                \\
                                &                                             & \multicolumn{1}{c|}{$1929-57i$}  & $(--++)$ & \multicolumn{1}{c|}{$1926-8i$} & $(--++)$
                                \\
                                &                                             & \multicolumn{1}{c|}{$1840-95i$}  & $(---+)$ & \multicolumn{1}{c|}{$1774-3i$} & $(---+)$\\ \hline
\multirow{3}{*}{$(-2,\frac12)$} & \multirow{3}{*}{$(1671, 1880, 2081, 2168)$} & \multicolumn{1}{c|}{$1600$}  & $(++++)$ & \multicolumn{1}{c|}{$1599$}    & $(++++)$  \\
                                &                                             & \multicolumn{1}{c|}{$1758-36i$} & $(-+++)$ & \multicolumn{1}{c|}{$1757-41i$}               &   $(-+++)$          \\
                                &                                             & \multicolumn{1}{c|}{$2041-94i$}  & $(--++)$ & \multicolumn{1}{c|}{$2023-45i$}               &     $(--++)$
                                \\
                                &                                             & \multicolumn{1}{c|}{$2008-104i$}  & $(---+)$ & \multicolumn{1}{c|}{$1981-178i$}               &     $(---+)$\\ \hline
$(-3,0)$                        & $(2029, 2220)$                              & \multicolumn{1}{c|}{$1860$}      & $(++)$   & \multicolumn{1}{c|}{$1860$} & $(++)$  \\ \hline\hline
\end{tabular}
\end{table*}

\section{Summary}
In this study, our primary focus is on investigating the interactions between the pseudoscalar meson octet and the baryon decuplet with $J^{P}=\frac32^+$, with the aim of identifying any potential resonances with $J^P=\frac32^-$. To achieve this, we have utilized slightly different subtraction constants compared to those used in Ref.~\cite{Sarkar:2004jh}. By doing so, we have identified several poles which potentially correspond to the resonances listed in the RPP~\cite{ParticleDataGroup:2022pth}, such as the $N(1857)$, $\Sigma(1670)$, $\Sigma(1910)$, $\Xi(1820)$, and $\Omega(2012)$. Furthermore, our findings confirm the two-pole structure of the $\Xi(1820)$, as pointed out in Ref.~\cite{Molina:2023uko}. Additionally, we place emphasis on the potential two-pole structure for the $\Sigma(1670)$.

Next, we examine the interaction between $\Phi \bm T$ in a finite volume. To regularize the two point Green's function, we introduce a form factor, which allows us to identify additional poles in these systems. These poles include several bound states that lie below the lowest threshold in each $(S,I)$ sector. Using UChPT as the underlying theory, we make predictions for the energy levels in finite volume with different lengths, denoted as $L$. These predicted energy levels are then fitted using an amplitude parameterized by the $K$-matrix, which satisfies flavor SU(3) symmetry. The utilization of SU(3) symmetry in the parameterization of the $K$-matrix helps reduce the number of parameters and improves the stability of the fit. The goodness of fit for the lowest four or five energy levels is satisfactory in each $(S,I)$ sector. By employing the parameters obtained from the best fits, we determine the pole positions of the $K$-matrix parameterized amplitude. We find that the $K$-matrix parametrization is able to accurately reproduce the poles of the underlying theory only when the pole is close to the physical region and its mass is not too far from the lowest relevant threshold. This suggests that when fitting the energy levels obtained from lattice study using the $K$-matrix parameterization, we should be mindful of the applicable energy region and be cautious when dealing with poles that are too high, as they may not be reliable. { Note that if one wants to extend the range of applicability of the method
discussed here, one has to include more terms (and poles) in the  $K$-matrix parametrization, which also increases the numebr of fit parameters. }

\begin{acknowledgments}
We express our gratitude to Anuvind Asokan, Feng-Kun Guo, Pan-Pan Shi and Meng-Na Tang for valuable discussions.  This work is supported by the
Deutsche Forschungsgemeinschaft (DFG, German Research Foundation) and the NSFC through the funds provided to the Sino-German Collaborative
Research Center TRR110 ``Symmetries and the Emergence of Structure in QCD'' (DFG Project ID 196253076 - TRR 110, NSFC Grant No. 12070131001), the Chinese Academy of Sciences (CAS) President's International Fellowship Initiative (PIFI) (Grant No. 2018DM0034) and Volkswagen Stiftung (Grant No. 93562).

\end{acknowledgments}

\bibliography{ref}

\begin{appendix}
    \section{SU(3) symmetric coupling structures}\label{app:CR}
    In this appendix we present the SU(3) symmetric coupling structures $C_{R}^{(S,I)}$.
    \begin{equation}
    \begin{aligned}
        C_{8}^{(0,\frac12)}&=\frac15 \left(
\begin{array}{cc}
 4 & -2 \\
 -2 & 1 \\
\end{array}
\right), \\
        C_{27}^{(0,\frac12)}&=\frac15 \left(
\begin{array}{cc}
 1 & 2 \\
 2 & 4 \\
\end{array}
\right).
\end{aligned}
\end{equation}

\begin{equation}
    \begin{aligned}
C_{8}^{(-1,0)}&=\frac 15 \left(
\begin{array}{cc}
 3 & \sqrt{6} \\
 \sqrt{6} & 2 \\
\end{array}
\right),\\
C_{27}^{(-1,0)}&=\frac 15\left(
\begin{array}{cc}
 2 & -\sqrt{6} \\
 -\sqrt{6} & 3 \\
\end{array}
\right).
\end{aligned}
\end{equation}

\begin{equation}
    \begin{aligned}
C_{8}^{(-1,1)}&=\frac15 \left(
\begin{array}{cccc}
 \frac{8}{3} & -2 \sqrt{\frac{2}{3}} & \frac{4}{3} & \frac{4}{3} \\
 -2 \sqrt{\frac{2}{3}} & 1 & -\sqrt{\frac{2}{3}} & -\sqrt{\frac{2}{3}} \\
 \frac{4}{3} & -\sqrt{\frac{2}{3}} & \frac{2}{3} & \frac{2}{3} \\
 \frac{4}{3} & -\sqrt{\frac{2}{3}} & \frac{2}{3} & \frac{2}{3} \\
\end{array}
\right),\\
C_{10}^{(-1,1)}&=\frac 13 \left(
\begin{array}{cccc}
 1 & -1 & 0 & 1 \\
 -1 & 1 & 0 & -1 \\
 0 & 0 & 0 & 0 \\
 1 & -1 & 0 & 1 \\
\end{array}
\right),\\
C_{27}^{(-1,1)}&=\frac{1}{20} \left(
\begin{array}{cccc}
 9 & 3 & 3 \sqrt{6} & -6 \\
 3 & 1 & \sqrt{6} & -2 \\
 3 \sqrt{6} & \sqrt{6} & 6 & -2 \sqrt{6} \\
 -6 & -2 & -2 \sqrt{6} & 4 \\
\end{array}
\right),\\
C_{35}^{(-1,1)}&=\frac{1}{12}\left(
\begin{array}{cccc}
 1 & -1 & -\sqrt{6} & -2 \\
 -1 & 1 & \sqrt{6} & 2 \\
 -\sqrt{6} & \sqrt{6} & 6 & 2 \sqrt{6} \\
 -2 & 2 & 2 \sqrt{6} & 4 \\
\end{array}
\right).
\end{aligned}
\end{equation}

\begin{equation}
    \begin{aligned}
C_{8}^{(-2,\frac12)}&=\frac 15 \left(
\begin{array}{cccc}
 1 & 1 & -1 & \sqrt{2} \\
 1 & 1 & -1 & \sqrt{2} \\
 -1 & -1 & 1 & -\sqrt{2} \\
 \sqrt{2} & \sqrt{2} & -\sqrt{2} & 2 \\
\end{array}
\right),\\
C_{10}^{(-2,\frac12)}&=\frac 18 \left(
\begin{array}{cccc}
 1 & -2 & 1 & \sqrt{2} \\
 -2 & 4 & -2 & -2 \sqrt{2} \\
 1 & -2 & 1 & \sqrt{2} \\
 \sqrt{2} & -2 \sqrt{2} & \sqrt{2} & 2 \\
\end{array}
\right),\\
C_{27}^{(-2,\frac12)}&=\frac{1}{80} \left(
\begin{array}{cccc}
 49 & 14 & 21 & -21 \sqrt{2} \\
 14 & 4 & 6 & -6 \sqrt{2} \\
 21 & 6 & 9 & -9 \sqrt{2} \\
 -21 \sqrt{2} & -6 \sqrt{2} & -9 \sqrt{2} & 18 \\
\end{array}
\right),\\
C_{35}^{(-2,\frac12)}&=\frac{1}{16} \left(
\begin{array}{cccc}
 1 & -2 & -3 & -\sqrt{2} \\
 -2 & 4 & 6 & 2 \sqrt{2} \\
 -3 & 6 & 9 & 3 \sqrt{2} \\
 -\sqrt{2} & 2 \sqrt{2} & 3 \sqrt{2} & 2 \\
\end{array}
\right).
\end{aligned}
\end{equation}

\begin{equation}
    \begin{aligned}
C_{10}^{(-3,0)}&=\frac 12 \left(
\begin{array}{cc}
 1 & 1 \\
 1 & 1 \\
\end{array}
\right),\\
C_{35}^{(-3,0)}&=\frac 12 \left(
\begin{array}{cc}
 1 & -1 \\
 -1 & 1 \\
\end{array}
\right).
\end{aligned}
\end{equation}

\begin{equation}
    \begin{aligned}
C_{10}^{(0,\frac32)}&=\frac 18 \left(
\begin{array}{ccc}
 5 & \sqrt{5} & -\sqrt{10} \\
 \sqrt{5} & 1 & -\sqrt{2} \\
 -\sqrt{10} & -\sqrt{2} & 2 \\
\end{array}
\right),\\
C_{27}^{(0,\frac32)}&=\frac{1}{16} \left(
\begin{array}{ccc}
 5 & -3 \sqrt{5} & \sqrt{10} \\
 -3 \sqrt{5} & 9 & -3 \sqrt{2} \\
 \sqrt{10} & -3 \sqrt{2} & 2 \\
\end{array}
\right),\\
C_{35}^{(0,\frac32)}&=\frac{1}{16} \left(
\begin{array}{ccc}
 1 & \sqrt{5} & \sqrt{10} \\
 \sqrt{5} & 5 & 5 \sqrt{2} \\
 \sqrt{10} & 5 \sqrt{2} & 10 \\
\end{array}
\right).
\end{aligned}
\end{equation}

\section{Parameters from best fits}\label{app:fitpars}
We collected here the parameters from the best fits for each $(S,I)$ sector.
\begin{equation}
    \begin{aligned}
    g_{8}^{(0,1/2)}&=769\ \rm MeV,\\
    m_{8}^{(0,1/2)}&=1343\ \rm MeV,\\
    c_{8}^{(0,1/2)}&=-4.1,\\
    c_{27}^{(0,1/2)}&=2.1.
\end{aligned}
\end{equation}

\begin{equation}
    \begin{aligned}
    g_8^{(-1,1)}&=48.7\ \rm MeV,\\
    m_8^{(-1,1)}&=1383\ \rm MeV,\\
    c_8^{(-1,1)}&=19.5,\\
    g_{10}^{(-1,1)}&=4275\ \rm MeV,\\
    m_{10}^{(-1,1)}&=1841\ \rm MeV,\\
    c_{10}^{(-1,1)}&=-44.3,\\
    c_{27}^{(-1,1)}&=-2.8,\\
    c_{35}^{(-1,1)}&=0.4.
\end{aligned}
\end{equation}

\begin{equation}
    \begin{aligned}
    g_8^{(-1,0)}&=984.5\ \rm MeV,\\
    m_8^{(-1,0)}&=1506\ \rm MeV,\\
    c_8^{(-1,0)}&=-7.2,\\
    c_{27}^{(-1,0)}&=2.1.
\end{aligned}
\end{equation}

\begin{equation}
    \begin{aligned}
    g_8^{(-2,1/2)}&=618\ \rm MeV,\\
    m_8^{(-2,1/2)}&=1599\ \rm MeV,\\
    c_8^{(-2,1.2)}&=-16.2,\\
    g_{10}^{(-2,1/2)}&=1725\ \rm MeV,\\
    m_{10}^{(-2,1/2)}&=1972\ \rm MeV,\\
    c_{10}^{(-2,1.2)}&=5.7,\\
    c_{27}^{(-2,1.2)}&=1.5,\\
    c_{35}^{(-2,1.2)}&=3.2.
\end{aligned}
\end{equation}

\begin{equation}
    \begin{aligned}
    g_8^{(0,3/2)}&=982\ \rm MeV,\\
    m_8^{(0,3/2)}&=1411\ \rm MeV,\\
    c_8^{(0,3/2)}&=-1.5,\\
    c_{27}^{(0,3/2)}&=-85.6,\\
    c_{35}^{(0,3/2)}&=-1.9.
\end{aligned}
\end{equation}

\begin{equation}
    \begin{aligned}
    g_{10}^{(-3,0)}&=1499\ \rm MeV,\\
    m_{10}^{(-3,0)}&=1860\ \rm MeV,\\
    c_{10}^{(-3,0)}&=-4.5,\\
    c_{35}^{(-3,0)}&=-3.6.
\end{aligned}
\end{equation}
\end{appendix}


\end{document}